\documentclass[%
 reprint,nofootinbib,
superscriptaddress,
 amsmath,amssymb,
 aps,
]{revtex4-2}

\usepackage{graphicx}
\usepackage{dcolumn}
\usepackage{bm}

\usepackage{xspace} 
\usepackage{bm}
\usepackage[colorlinks = true,citecolor=red]{hyperref}
\usepackage{float}
\usepackage[caption=false]{subfig}
\usepackage[capitalise,noabbrev]{cleveref}
\usepackage{soul}

\newcommand{\LCDM}{$\Lambda$CDM\xspace}
\newcommand{\aap}{A\&A}
\newcommand{\mnras}{MNRAS}
\newcommand{\jcap}{J. Cosmology Astropart. Phys}

\newif\ifhighlightchanges
\highlightchangesfalse 
\newcommand{\change}[1]{\ifhighlightchanges{\color{red} #1}\else #1 \fi}

\begin{document}

\title{Constraining Quantum Initial Conditions Before Inflation}

\author{T. Gessey-Jones}
    \email{tg400@cam.ac.uk}
    \affiliation{Astrophysics Group, Cavendish Laboratory, J.J. Thomson Avenue, Cambridge, CB3 0HE, UK}
\author{W. J. Handley}
    \email{wh260@mrao.cam.ac.uk}
    \affiliation{Astrophysics Group, Cavendish Laboratory, J.J. Thomson Avenue, Cambridge, CB3 0HE, UK}
    \affiliation{Kavli Institute for Cosmology, Madingley Road, Cambridge, CB3 0HA, UK}
    \affiliation{Gonville \& Caius College, Trinity Street, Cambridge, CB2 1TA, UK}

\begin{abstract}
We theoretically and observationally investigate different choices of initial conditions for the primordial mode function that are imposed during an epoch preceding inflation. 
By deriving predictions for the observables resulting from several alternate quantum vacuum prescriptions we show some choices of vacua are theoretically observationally distinguishable from others. 
Comparing these predictions to the Planck 2018 observations via a Bayesian analysis shows no significant evidence to favour any of the quantum vacuum prescriptions over the others. 
In addition we consider frozen initial conditions, representing a white-noise initial state at the big-bang singularity. 
Under certain assumptions the cosmological concordance model and frozen initial conditions are found to produce identical predictions for the cosmic microwave background anisotropies. 
Frozen initial conditions may thus provide an alternative theoretic paradigm to explain observations that were previously understood in terms of the inflation of a quantum vacuum. 
\end{abstract}

\maketitle

\section{\label{sec:intro}Introduction} 
Since the discovery in 1998 that the expansion of the universe is accelerating, the cosmological concordance model (\LCDM) has become the standard model of cosmology~\cite{Riess_1998,Douglas_2018}. Subsequent experimental results have continued to be in good agreement with its predictions, with the most recent results from the Planck satellite finding ``no compelling evidence for extensions to the base-\LCDM model''~\cite{PLANCK_2018_VI}. Despite the success of \LCDM there are still small features in the cosmic microwave background (CMB) power spectra that remain unexplained~\cite{PLANCK_2018_VII}. 

A key component of \LCDM is the idea that the early universe was nearly homogeneous, but did have small deviations about this mean~\cite{Mukhanov_Cosmology}. These primordial fluctuations formed just after the big bang and then subsequently grew under their own gravity. The growing fluctuations later caused the small anisotropies seen in the CMB and eventually led to the large scale structure of the universe. In \LCDM it is assumed that the power spectrum of these primordial fluctuations is a power law.

An inflationary epoch in the very early universe can naturally lead to such a power law spectrum developing from a quantum vacuum~\cite{Baumann_2009}.  During inflation the quantum fluctuations inherent in the vacuum are inflated to macroscopic scales by the accelerating growth of the universe. There remains some ambiguity in this model as in curved or rapidly evolving spacetime there is not a unique vacuum state~\cite{Fulling_1989,Birrell_1982}. As a result many  vacua~\cite{Handley_2016b} have been proposed as potential initial states of the universe, each with their own merits~\cite{Agocs_2020b}. 

More recent work~\cite{Handley_2014,Handley_2016,Hergt_2019,2019PhRvD.100b3501H} has investigated an alternative paradigm where the inflationary period is preceded by a period of kinetic dominance. In this paradigm the universe's initial state would be set during this earlier kinetic dominance phase rather than during inflation. This approach is fairly similar to that adopted in the ``just enough inflation'' scenario~\cite{Ramirez_2012a,Ramirez_2012b} where the initial state is set at the beginning of fast-roll inflation. 

In this paper we investigate the effect that setting the universe's initial state during a kinetic dominance era has on the primordial power spectrum and experimental observables. For the initial state we consider a range of alternative initial vacua with the goal of determining if the choice of quantum vacuum may be observationally distinguishable. In addition we investigate an alternative to the paradigm of setting vacuum initial conditions, frozen initial conditions~\cite{Haddadin_2018b}, which instead represent a white-noise initial perturbation state at the big-bang singularity. By comparing the observable predictions to the Planck 2018 experimental results~\cite{PLANCK_2018_V}, we then aim to determine if any of the proposed initial conditions can provide an explanation for the features in the CMB spectra not explained by \LCDM. Throughout we adopt a generic approach that does not depend on the exact form of the inflaton potential, allowing us to disentangle the effects of initial conditions from the effects of the choice of potential~\cite{Contaldi_2003}.

\Cref{sec:thry} details the theoretical background to this paper, including the specification of all initial conditions that are considered. From \cref{sec:aic} onward we develop approximate analytic expressions for the primordial power spectrum resulting from each of the various initial conditions. Using these in \cref{sec:obs} the corresponding CMB anisotropy spectra are calculated and compared. Following which in \cref{sec:bayes} we use Bayes' factor and the Planck likelihoods~\cite{PLANCK_2018_V} to determine which model is best supported by the observational data. Finally in \cref{sec:conc} we present our conclusions.
\vfill

\section{\label{sec:thry}Theoretical Background}
The results and derivations quoted in this section have been taken from~\cite{Baumann_2009,Handley_2016b,HEL_GR} unless otherwise stated. All equations are given in natural units where $c = \hbar = 8 \pi G = 1$. Cosmic time derivatives will be denoted by overdots ($\dot{a}$) and conformal time derivatives by dashes ($a'$).

\subsection{\label{ssec:be}Background Equations}
Consider a universe containing only a canonical scalar field ($\phi$). Such a universe can be described by the Einstein-Hilbert action
\begin{equation}
    S = \int d^4x \sqrt{|g|} \left( \frac{1}{2}R + \frac{1}{2}\nabla^\mu \phi \nabla_\mu \phi - V(\phi) \right).
    \label{eq:EH_action}
\end{equation}
Where $g_{\mu \nu}$ is the metric, $R$ the Ricci scalar and $V(\phi)$ the potential energy of the $\phi$ scalar field. 

Initially let us assume said universe is homogeneous, isotropic and spatially flat. Imposing these constraints on $g_{\mu \nu}$ and $\phi$, then extremizing \cref{eq:EH_action} gives
\begin{align}
    ds^2 = dt^2 - a^2\delta_{ij}dx_i dx_j, 
    \label{eq:FRW_flat}
\\
    \dot{H} + H^2 = -\frac{1}{3} \left(\dot{\phi}^2 - V(\phi)\right) ,
    \label{eq:BE1}
\\
    \ddot{\phi} + 3H\dot{\phi} + \frac{dV}{d\phi}= 0.
    \label{eq:BE2}
\end{align}
Where $a$ is the scale factor of the universe and $H$ the Hubble parameter, defined to be $H \equiv \dot{a}/a$. It is often convenient to re-express \cref{eq:BE1} as its first integral
\begin{equation}
    H^2 = \frac{1}{3} \left(\frac{1}{2}\dot{\phi}^2 + V(\phi)\right).
    \label{eq:BE3}
\end{equation}
Once initial conditions on $\phi$, $\dot{\phi}$, and $a$ are imposed, any two of \cref{eq:BE1,eq:BE2,eq:BE3} are sufficient to fully specify the evolution of this flat, homogeneous, isotropic universe. Collectively these three equations are referred to as the background equations.

\subsection{\label{ssec:ms}Mukhanov-Sasaki Equations}

CMB observations suggest that the early universe was very close to homogeneous, isotropic and flat with only small deviations from the ideal case. As these deviations were small it should be sufficient to limit ourselves to considering linear perturbations of the scalar field and metric about their homogeneous background values 
\begin{align}
    \phi(t,\bm{x}) = & \textrm{ } \bar{\phi}(t) + \delta \phi(t,\bm{x}) ,
    \label{eq:PE_phi}
\\
    ds^2 = & (1+2\Phi)dt^2 + 2a\left(\partial_i B - S_i\right)dx_i dt  \nonumber \\&-a^2\left(1-2\Psi\right)\delta_{ij}dx_i dx_j  \nonumber \\ &-a^2\left(2\partial_i\partial_j E + 2 \partial_{(i} F_{j)} + h_{ij}\right)dx_i dx_j.
    \label{eq:PE_met}
\end{align}
In \cref{eq:PE_met} the metric perturbation has been split into several scalar, vector and tensor components (SVT decomposition). For linear perturbations these three types of component do not dynamically mix, hence allowing the vector and tensor components to be safely neglected in the following derivations.

Cosmological perturbation theory is plagued by the complication that the definitions of perturbations and background are not unique, as a coordinate transformation can change a density perturbation into a metric perturbation and vice versa. To avoid the issue it is usual to define gauge invariant combinations of the metric and field perturbations that do not change under coordinate transformations. The gauge invariant quantity of interest for calculating the primordial power spectrum is the comoving curvature perturbation $\mathcal{R}$, which can be interpreted as the curvature of the spatial hypersurfaces of constant $\phi$, and is defined to be
\begin{equation}
    \mathcal{R} \equiv \Psi - \frac{H}{\dot{\phi}}\delta{\phi}. 
    \label{eqn:R_def}
\end{equation}

Substituting \cref{eq:PE_phi,eq:PE_met,eqn:R_def} into \cref{eq:EH_action} gives us a general action for $\mathcal{R}$. As we are considering small perturbations $\mathcal{R}$ is small, so let us only keep the lowest order terms in the action
\begin{equation}
    S_{\mathcal{R}} = \frac{1}{2}\int d^4x\; a^3\left(\frac{\dot{\phi}}{H}\right)^2 \left(\dot{\mathcal{R}}^2 - \left(\frac{\partial_i \mathcal{R}}{a}\right)^2\right).
    \label{eqn:S_R}
\end{equation}
Which can be simplified by making a change of variable to the Mukhanov variable, $v \equiv z \mathcal{R}$, along with a coordinate transformation to conformal time, $\eta \equiv \int {dt}/{a}$
\begin{equation}
    S_{v} = \frac{1}{2}\int d\eta d^3x  \left(v'^2 - (\partial_i v)^2 + \frac{z''}{z}v^2\right),
    \label{eqn:S_v}
\end{equation}
where for convenience $z$ is defined to be $z \equiv {a\dot{\phi}}/{H}$.

Finally, Fourier transforming in the spatial coordinates followed by extremizing the resulting action gives the Mukhanov-Sasaki equation
\begin{equation}
    v_k'' + \left(k^2 - \frac{z''}{z}\right)v_k = 0. 
    \label{eqn:MS}
\end{equation}
This equation implicitly describes the evolution of comoving curvature perturbations with a comoving wavevector of magnitude $k$. Its behaviour depends strongly on the state of the homogeneous background universe via the $z''/z$ term.

\subsection{\label{ssec:kd_inf}Kinetic Dominance and Inflation}
\begin{figure}[t]
\includegraphics{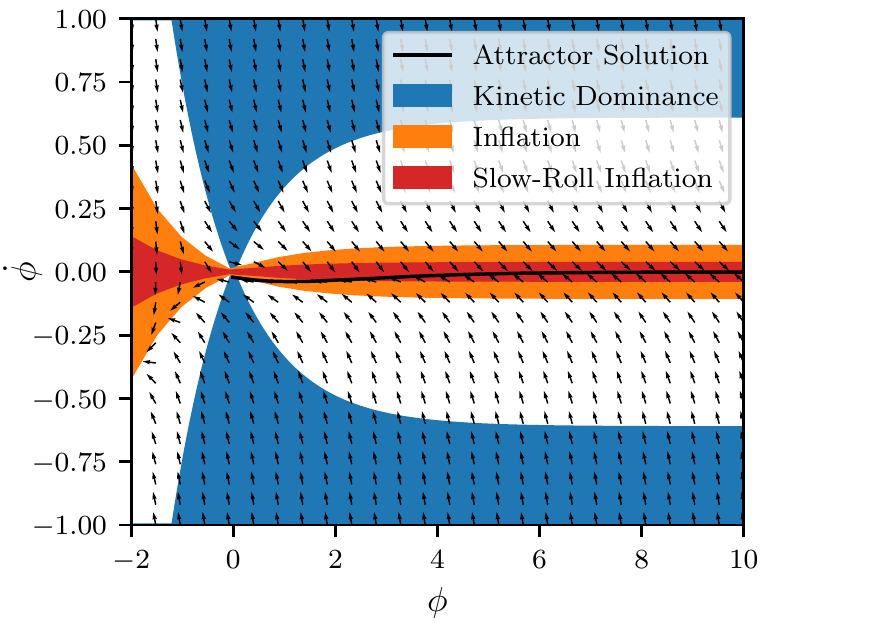}
\caption{\label{fig:phase_plot} Phase portrait of scalar field inflation with a Starobinsky potential $V(\phi) = \Lambda^4\left(1 - \exp\left[-\sqrt{2/3}\phi\right] \right)^2$~\cite{Starobinsky_1980} where \mbox{$\Lambda^2 = 0.1$}. Arrows denote the direction of system evolution in phase space. Kinetic dominance \mbox{($w_\phi > 0.9$)}, inflation \mbox{($w_\phi < -\frac{1}{3}$)}, and slow-roll inflation \mbox{($w_\phi <-0.9$)} regimes are highlighted. The plot visualises the evolution of possible universes under this model, universes beginning with $\phi > 0$ typically start in kinetic dominance and then move onto the attractor solution inside slow-roll inflation. Plot is based upon Figure 3 in~\cite{Hergt_2019}.}
\end{figure}

Following the recent work in~\cite{Handley_2014,Handley_2016,Hergt_2019} which gives support to the idea of a kinetic dominance era preceding inflation, we shall suppose the universe exits the Planck era into a state of kinetic dominance, where $\frac{1}{2}\dot{\phi}^2 \gg V(\phi)$. In such a phase $\phi$ changes rapidly (see \cref{fig:phase_plot}) but decelerates due to the high effective friction, and so eventually kinetic dominance ends. After a short period passing through a fast-roll inflation stage these models then tend to enter a period of slow-roll inflation, $\frac{1}{2}\dot{\phi}^2 \ll V(\phi)$. In slow-roll inflation the field changes very slowly, causing $H$ to remain almost constant, leading to near exponential growth of the universe's scale factor. However $\dot{\phi}$ is not exactly zero and so $H$ slowly decreases as $\phi$ tends to the minimum of its potential. Inflation finally ends when the field $\phi$ approaches the bottom of the inflationary potential and typically begins to oscillate around the minimum.

\begin{figure}[t]
\includegraphics{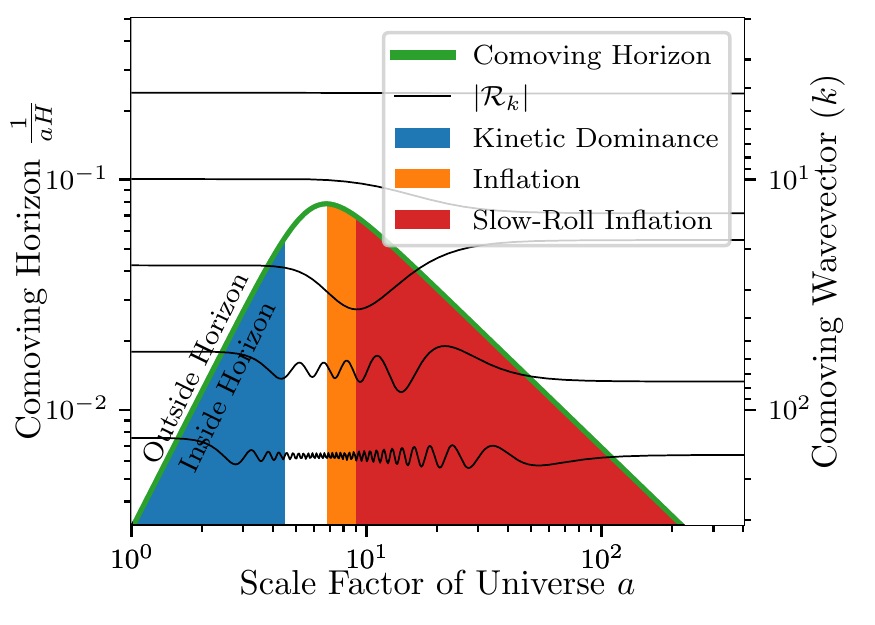}
\caption{\label{fig:mode_evo} Evolution of the comoving horizon and $\mathcal{R}_k$ modes. For illustrative purposes the $R_k$ modes are shifted in the vertical direction to be about the corresponding value of $k$. The modes oscillate when they enter the horizon during kinetic dominance, and then freeze out again once they leave during inflation. Some modes at low $k$ however always remain far above the horizon and so stay frozen throughout. Figure is based upon Figure 1 in~\cite{Haddadin_2018b}. }
\end{figure}

From \cref{fig:mode_evo} we can see the typical behaviour of the perturbations during kinetic dominance and inflation. When a perturbation mode is outside of the comoving horizon, $1 /aH  < 1 / k$, it freezes-out and stops evolving. Conceptually, freezing-out can be thought of as the expansion of the universe being too rapid for that perturbation mode to be in causal contact with itself, and so it cannot evolve. Conversely inside the horizon the perturbation modes rapidly oscillate. Note as the comoving horizon has a maximum, the modes corresponding to the largest length-scales never enter the horizon and so stay frozen throughout. 

By the end of inflation the comoving horizon has shrunk greatly, leaving all the modes of observational interest outside the horizon and hence frozen. These frozen modes form the scalar primordial power spectrum, defined to be
\begin{equation}
    \mathcal{P}_\mathcal{R}(k) = \lim_{k \ll aH} \frac{k^3}{2\pi^2}\left|\mathcal{R}_k\right|^2.
    \label{eqn:pps}
\end{equation}

After inflation ends the comoving horizon starts expanding again. As a result the frozen modes eventually re-enter the comoving horizon and unfreeze. Once unfrozen the modes begin growing due to their own gravity and in time lead to the CMB anisotropies and large scale structure which can be experimentally observed.

\subsection{\label{ssec:ic}Initial Conditions}
To summarise, deriving the primordial power spectrum amounts to solving the background and Mukhanov-Sasaki equations up until the modes of interest freeze-out. To fully solve the equations requires the specification of initial conditions on the perturbation modes. In our model the earliest era in the universe's history is the kinetic dominance epoch, thus it seems natural to specify the modes' initial conditions during said epoch. 

For standard inflationary models the perturbation modes' initial conditions are typically chosen to be a quantum vacuum set far back in the inflationary epoch. While in a rapidly evolving spacetime there is no unique vacuum state, many of the proposed quantum vacua tend to a Bunch-Davies vacuum~\cite{Baumann_2009} for modes well within the comoving horizon. Since in classical inflation the comoving horizon expands endlessly as we approach the initial singularity, by imposing the initial conditions far enough back into the inflationary epoch all observable scales can be set deep within the comoving horizon. Consequently the choice between such initial conditions does not effect observables~\cite{Birrell_1982,Fulling_1989}. However, if the initial state of observable modes are set at a time when the mode is not sufficiently deep within the comoving horizon, the choice of quantum vacuum can have an impact on observables. This was illustrated by \citet{Sriramkumar_2005}, who demonstrated how different quantum initial states could in general lead to primordial power spectra identical to those predicted by phenomenological models of trans-Planckian physics.   

In a model of the universe with a kinetic dominance epoch preceding inflation the comoving horizon now has a maximum. As a result it may not be possible to set the mode initial conditions at a time when all observable scales are deep within the comoving horizon. Consequently we expect the choice of quantum vacuum may effect the resulting observables which presents us the problem of which is the correct quantum vacuum to impose. Conversely, if the choice of quantum vacuum does impact the observables then said observables can in theory be used to constrain the choice of quantum vacuum. Hence in this paper we aim to investigate this prospects of using cosmological observations to distinguish potential quantum vacuums. 

We shall consider the following mode initial conditions: Bunch-Davies vacuum (BD)~\cite{Baumann_2009}, Hamiltonian Diagonalisation (HD)~\cite{Handley_2016b}, Renormalized Stress Energy Tensor (RSET)~\cite{Handley_2016b}, Right Handed Mode (RHM)~\cite{Contaldi_2003}, and Frozen Initial Conditions (FIC)~\cite{Haddadin_2018b}. The mathematical definitions of each of these are given in \cref{tab:ICs}. 

\renewcommand{\arraystretch}{1.8}
\begin{table}[t]
\caption{\label{tab:ICs}%
Table of the mathematical definitions of all types of initial conditions considered in this paper. Where $\textrm{H}^{(2)}$ denotes a Hankel function of the second kind.}
\begin{ruledtabular}
\begin{tabular}{ccc}

\textrm{Initial Condition}&
\multicolumn{2}{c}{\textrm{Definition}}\\

\colrule
BD & $v_k = \frac{1}{\sqrt{2k}}$ & $v_k' = -i k v_k$\\
HD & $v_k = \frac{1}{\sqrt{2k}}$ & $v_k' = -i \sqrt{k^2 - \frac{z''}{z}} v_k$\\
RSET & $v_k = \frac{1}{\sqrt{2k}}$ & $v_k' = (-ik + \frac{z'}{z}) v_k$\\
RHM & \multicolumn{2}{c}{$v_k(\eta) = \sqrt{\frac{\pi}{8k_t}}\sqrt{1+ 2k_t \eta} \textrm{H}_0^{(2)}\left[k\eta + \frac{k}{2k_t}\right]$}\\
FIC & $R_k = R^{(0)} = \textrm{const.}$ & $R_k' = 0$\\
\end{tabular}
\end{ruledtabular}
\end{table}

BD, HD, RSET, and RHM, are all derived from previously suggested initial quantum vacua. Note, this list is far from exhaustive and many other vacua have been proposed~\cite{Danielsson_2002,Chernikov_1968,Tagirov_1972,Allen_1985,Mottola_1984}. There remains much debate as to which of the proposed vacua is the correct one to set. Detailed discussions of the theoretical merits and issues of some of these different choices can be found in~\cite{Handley_2016b,Agocs_2020b}.

FIC are distinct from the other initial conditions we shall consider as they do not represent a quantum vacuum state. Hence setting them goes against the common assumption that the perturbations were initially in some form of quantum vacuum. Instead FIC are a white-noise state where all modes start with equal magnitude and zero velocity. They were introduced and are briefly explored numerically in~\cite{Haddadin_2018b}.

\subsection{\label{ssec:apps_meth}Approximating the Primordial Power Spectra}
The background and Mukhanov-Sasaki equations do not have general analytic solutions. However, approximate solutions for them are known in some regimes, which can then be used to find approximations of the resulting primordial power spectrum.

One such primordial power spectrum approximation method was proposed by \citet{Contaldi_2003}, which we shall refer to as the Contaldi approximation. This method has the notable advantage of not depending on the potential of the scalar field, allowing us to form fairly general conclusions about the effects of perturbation initial conditions. Recently this method has been extended by \citet{Thavanesan_2020} to curved universes. However, in this paper we consider flat universes only and so the method of \citet{Contaldi_2003} is sufficient for our purposes. Broadening our investigation to consider open and closed universes is left for future studies~\cite{Gessey-Jones_2020c}.

\begin{figure}[t]
\includegraphics{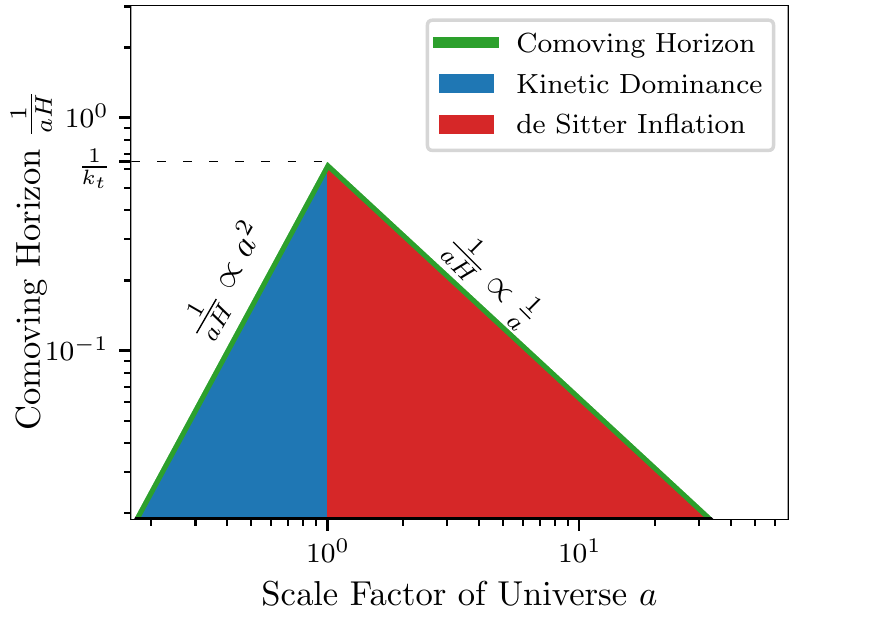}
\caption{\label{fig:contaldi_approx} Comoving horizon in the Contaldi approximation. Note how, unlike in \cref{fig:mode_evo}, the behaviour of the comoving horizon changes discontinuously at a well defined transition point. At this transition point the the comoving horizon takes its maximum value $1 / k_t$.}
\end{figure}

The key to the approximation is modelling the background universe as being in a kinetic dominance phase followed by a de Sitter inflation phase with an instantaneous transition between them (illustrated in \cref{fig:contaldi_approx}). At the transition $a$ and $H$ are matched to ensure continuity of the background universe.

\citet{Contaldi_2003} showed that in both the kinetic dominance and de Sitter inflation regime the Mukhanov-Sasaki equations have analytic solutions. For kinetic dominance the solution is
\begin{multline} 
    v_k(\eta) = \sqrt{k\eta + \frac{k}{2k_t}} \bigg( A_k \textrm{H}_0^{(1)}\left[k\eta + \frac{k}{2k_t}\right] \\ +  B_k \textrm{H}_0^{(2)}\left[k\eta + \frac{k}{2k_t} \right] \bigg).
    \label{eqn:vk_KD}
\end{multline}
$\textrm{H}^{(1/2)}_i$ here refer to the Hankel functions, and \mbox{$k_t = (aH)_{\textrm{trans}}$} is the reciprocal of the size of the comoving horizon at the transition, which is left as a free parameter\footnote{\citet{Contaldi_2003} take $a = 1$ at the transition.}. Whereas the solution in de Sitter inflation is
\begin{multline} 
    v_k(\eta) = C_k e^{-i\left(k\eta - \frac{k}{k_t}\right)}\left(1- \frac{i}{k\eta - \frac{k}{k_t}}\right) \\+ D_k e^{+i\left(k\eta - \frac{k}{k_t}\right)}\left(1+ \frac{i}{k\eta - \frac{k}{k_t}}\right).
    \label{eqn:vk_dS}
\end{multline}
For mathematical convenience the transition is taken to occur at $\eta = 0$. $A_k$, $B_k$, $C_k$, and $D_k$ are integration constants. By requiring continuity of $v_k$ and $v_k'$ across the transition the number of free integration constants is reduced by two, and the final two integration constants are fixed by the initial conditions.

Furthermore from \cref{eqn:vk_dS} \citet{Contaldi_2003} derived the primordial power spectrum under their approximation to be
\begin{equation} 
    \mathcal{P}_\mathcal{R}(k) = \frac{k_t^2 k}{4 \varepsilon a_t^2 \pi^2}\left|C_k - D_k\right|^2,
    \label{eqn:general_pps}
\end{equation}
where $\varepsilon$ is one of the slow-roll parameters~\cite{Baumann_2009}, and $a_t$ the scale factor at the kinetic dominance to inflation transition. 

In their original paper \citet{Contaldi_2003} consider RHM initial conditions corresponding to $A_k = 0$, \mbox{$B_k = \sqrt{\pi /4k}$}. Matching $v_k$ and $v_k'$ at the transition then gives
\begin{equation}
\label{eqn:cd_rhm}
\begin{split}
    C_k^{\textrm{(RHM)}} = 
    \sqrt{\frac{\pi}{32k_t}}&e^{-i\frac{k}{k_t}}\bigg(\textrm{H}_0^{(2)}\left[\frac{k}{2k_t}\right] \\ & - \left(\frac{k_t}{k} + i\right)\textrm{H}_1^{(2)}\left[\frac{k}{2k_t}\right]\bigg), 
\\
    D_k^{\textrm{(RHM)}} =  \sqrt{\frac{\pi}{32k_t}}&e^{+i\frac{k}{k_t}}\bigg(\textrm{H}_0^{(2)}\left[\frac{k}{2k_t}\right] \\& - \left(\frac{k_t}{k} - i\right)\textrm{H}_1^{(2)}\left[\frac{k}{2k_t}\right]\bigg),
\end{split}
\end{equation}
and hence an explicit form for the primordial power spectrum via \cref{eqn:general_pps}.

As outlined above the Contaldi approximation is predicated on the assumption that the Mukhanov-Sasaki equations can be used to describe the universe's perturbations in both the kinetic dominance and inflationary regimes, which in turn would require the perturbations in the universe at those times to be small. In their original paper \citet{Contaldi_2003} assume this is true while acknowledging that it may require a previous period of inflation to justify small perturbations in the kinetic dominance regime. However, it has subsequently be shown in~\cite{Hergt_2020,Hergt_2021b} that a kinetic dominance epoch, like inflation, acts to homogenise the universe. Hence, the use of the Mukhanov-Sasaki equations to describe perturbations at the end of the kinetic dominance era can now be somewhat justified by arguing the early stages of kinetic dominance act to make any perturbations small.

We shall now us this same method to derive the primordial power spectra resulting from the other initial conditions listed in \cref{tab:ICs}.

\section{\label{sec:aic}Primordial Power Spectra}

\subsection{\label{ssec:apps_tr}Vacuum Initial Conditions}
First let us consider imposing BD, HD and RSET initial conditions during the kinetic dominance epoch. We shall set these initial conditions at the kinetic dominance side of the kinetic dominance to inflation transition, where the comoving horizon is at its greatest extent. Doing so corresponds most closely to the idea of setting vacuum initial conditions very deep in the inflationary epoch which is commonly employed in models that do not consider a kinetic dominance phase. 

\begin{figure}[t]
\includegraphics{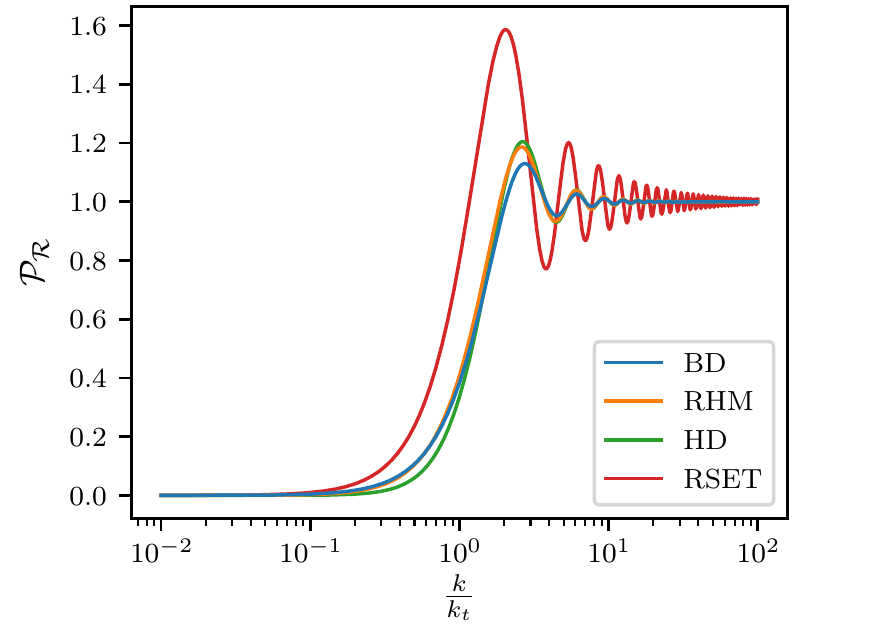}
\caption{\label{fig:pps_qic} Approximations of the primordial power spectra produced by BD, RHM~\cite{Contaldi_2003}, HD and RSET initial conditions. BD, RHM, and HD all produce very similar spectra, with a cutoff when $k<2k_t$, and a flat spectra when $k>10k_t$, with a small amount of oscillation in between. In RSET however the oscillations are far more prominent and die away far more slowly.}
\end{figure}

Hence the initial conditions directly give us $v_k(0-)$ and $v'_k(0-)$. Here we make the distinction between $0-$ and $0+$ clear since for HD and RSET initial conditions some of the background-dependent terms in their definitions change abruptly at the kinetic dominance to inflation transition. As we wish to study initial conditions imposed during the kinetic dominance epoch we only consider the $0-$ case. Imposing initial conditions at $0+$ instead would be equivalent to ignoring the presence of the kinetic dominance epoch in our calculations. The $v_k(0-)$ and $v'_k(0-)$ calculated from the initial conditions can then be matched onto the de Sitter solution for $v_k$ to derive the following results: for Bunch-Davies
\begin{align}
\begin{split} \label{eqn:cd_bd}
    C_k^{\textrm{(BD)}} = & \sqrt{\frac{1}{8k}}e^{-i\frac{k}{k_t}}\bigg(\left(\frac{k_t}{k}\right)^2 + 2i\left(\frac{k_t}{k}\right) - 2\bigg),\\
    D_k^{\textrm{(BD)}} = & \sqrt{\frac{1}{8k}}e^{+i\frac{k}{k_t}}\left(\frac{k_t}{k}\right)^2 ,
\end{split}
\end{align}
for hamiltonian diagonalisation 
\begin{align}
\begin{split}\label{eqn:cd_hd}
    C_k^{\textrm{(HD)}} = & \sqrt{\frac{1}{8\omega_k}}e^{-i\frac{k}{k_t}}\bigg(\left(\frac{k_t}{k}\right)^2\\ & + i\left(1 + \frac{\omega_k}{k}\right)\frac{k_t}{k} -  \left(1 + \frac{\omega_k}{k}\right)\bigg),\\
    D_k^{\textrm{(HD)}} = & \sqrt{\frac{1}{8\omega_k}}e^{+i\frac{k}{k_t}}\bigg(\left(\frac{k_t}{k}\right)^2 \\&- i\left(1 - \frac{\omega_k}{k}\right)\frac{k_t}{k} -  \left(1 - \frac{\omega_k}{k}\right)\bigg),
\end{split}
\end{align}
where $\omega_k^2 = k^2 + k_t^2$, and finally for renormalised stress energy tensor
\begin{align}
\begin{split}\label{eqn:cd_rset}
    C_k^{\textrm{(RSET)}} &=  \sqrt{\frac{1}{8k}}e^{-i\frac{k}{k_t}}\left(i\frac{k_t}{k} - 2\right),\\
    D_k^{\textrm{(RSET)}} &=  \sqrt{\frac{1}{8k}}e^{+i\frac{k}{k_t}}\left(i\frac{k_t}{k} \right).
\end{split}
\end{align}
Substituting these into \cref{eqn:general_pps} gives the approximate primordial power spectra resulting from BD, HD and RSET initial conditions, depicted in \cref{fig:pps_qic} along with the results \citet{Contaldi_2003} found for RHM initial conditions. 

From \cref{fig:pps_qic} we can see that $\mathcal{P}_\mathcal{R}$ for BD, RHM, and HD are all very similar. They all show a steep low $k$ cutoff below $k \approx 2k_t$ and at high $k$ a flat primordial power spectrum, with an intermediate region that shows some small oscillations. RSET however is noticeably different, the low $k$ cutoff being at a lower $k$ value and the oscillations about the high $k$ plateau are much larger and decay far more slowly. The vacuum initial conditions also have different behaviours below their low $k$ cutoff. In that region we find they obey power laws: BD and RSET $\propto k^2$, HD $\propto k^3$, and RHM $\propto k^3 (\log(k))^2$. Unfortunately for the values of $k_t$ we find to be supported by the Planck likelihoods (see \cref{sec:bayes}), this low $k$ region corresponds to scales that are too large to currently be observed. 

Hence the key features in these primordial power spectra, which may allow for experimentally distinguishing between vacuum initial conditions, are the position of the low $k$ cutoff and the size of any oscillations in the intermediate $k$ region. Therefore qualitatively it seems RSET may be observationally distinguishable from the other three, this idea has been previously discussed in~\cite{Handley_2016b,Agocs_2020b}. 

It could be argued that it would be more natural to impose perturbation mode initial conditions at the start of the universe, i.e. in  our model at the beginning of the kinetic dominance epoch. However from \cref{eqn:vk_KD} we find as \mbox{$\eta \to -\frac{1}{2k_t}$, $v_k \to 0$}. Therefore none of BD, HD or RSET initial conditions can be set at the singularity. Furthermore setting these initial conditions at the singularity would seem to be in contrast to their primary initial motivation of ensuring sub-comoving horizon modes are quantized, since at the singularity all modes are in the super-horizon limit. Away from the singularity $v_k$ is not automatically zero, so the vacuum initial conditions could mathematically be imposed at an arbitrary point in the kinetic dominance epoch, effectively introducing another free parameter into our model, corresponding to when during kinetic dominance the initial conditions were set. Investigating the impact of introducing this additional degree of freedom into the primordial power spectrum, and discussing the validity of using vacuum initial conditions away from the comoving horizon maximum, is left for future work~\cite{Gessey-Jones_2020c}.

\subsection{\label{ssec:apps_bb}Frozen Initial Conditions}
As described in \cref{ssec:kd_inf} perturbation modes outside the horizon freeze out and stop evolving, this is equally true in kinetic dominance as it is during inflation. Therefore we might expect that any modes that start outside the comoving horizon should begin frozen, or at least would rapidly freeze if they did start with some velocity. This motivates the idea of imposing frozen initial conditions deep within the kinetic dominance epoch, where all modes of observational interest are outside of the horizon and so have zero velocity. It is not immediately obvious what the initial magnitude of these modes should be, following \cite{Haddadin_2018b} we assume the $R_k$ modes all begin with an equal magnitude. It may be possible to motivate such a white noise state ab initio via bouncing cosmology, though to do so is outside the scope of this paper~\cite{Thavanesan_2021}. 

A further motivation for considering FIC is that while BD, HD and RSET initial conditions cannot be set at the big bang singularity, FIC can. From \cref{eqn:vk_KD} we find that during kinetic dominance
\begin{equation} 
\begin{split}
\mathcal{R}_k(\eta) & = \sqrt{\frac{k}{12k_t}}\bigg(A_k \textrm{H}_0^{(1)}\left[k\eta + \frac{k}{2k_t}\right] \\ & +  B_k \textrm{H}_0^{(2)}\left[k\eta + \frac{k}{2k_t}\right]\bigg).
\end{split}
\end{equation}
Re-expressing the above in terms of Bessel functions of the first ($\textrm{J}$) and second ($\textrm{Y}$) kinds gives
\begin{equation} 
\begin{split}
\mathcal{R}_k(\eta) & = \sqrt{\frac{k}{12k_t}}\bigg(\big(A_k + B_k\big) \textrm{J}_0\left[k\eta + \frac{k}{2k_t}\right] \\ & +  i\big(A_k-B_k\big) \textrm{Y}_0\left[k\eta + \frac{k}{2k_t}\right]\bigg).
\end{split}
\end{equation}
Taking the limit towards the singularity \mbox{($\eta \to -\frac{1}{2k_t}$)} $\mathcal{R}_k$ remains finite providing $A_k = B_k$.  Thus taking $A_k = B_k$ we find
\begin{align} 
&\mathcal{R}_k(\eta)  = \sqrt{\frac{k}{3k_t}} A_k \textrm{J}_0\left[k\eta + \frac{k}{2k_t}\right], \\
&\mathcal{R}_k'(\eta) = -\sqrt{\frac{k^3}{3k_t}} A_k \textrm{J}_1\left[k\eta + \frac{k}{2k_t}\right], \\
&\mathcal{R}_k'\left(-\frac{1}{2k_t}\right)  = 0.
\end{align}
Hence the FIC $\mathcal{R}'_k$ initial condition is automatically satisfied once we impose the condition $R_k = R^{(0)}$ by setting
\begin{equation}
    A_k = B_k = \sqrt{\frac{3k_t}{k}}R^{(0)}.
\end{equation}

Imposing initial conditions at the singularity is appealing as it removes any arbitrariness in when to impose the initial conditions on the perturbation modes. Therefore a priori FIC could be motivated by their compatibility with being set at the classical singularity unlike the vacuum initial conditions we have considered. Investigating if taking into account quantum gravity effects changes this result is left for future study.

Given we now know $A_k$ and $B_k$ for FIC, by matching $v_k$ and $v_k'$ at the transition we can arrive at
\begin{equation}
\begin{split}
    C_k^{\textrm{(FIC)}} = & \sqrt{\frac{3}{2}}R^{(0)}e^{-i\frac{k}{k_t}}\bigg(\textrm{J}_0\left[\frac{k}{2k_t}\right] - \left(\frac{k_t}{k} + i\right)\textrm{J}_1\left[\frac{k}{2k_t}\right]\bigg),
\\
    D_k^{\textrm{(FIC)}} = & \sqrt{\frac{3}{2}}R^{(0)}e^{+i\frac{k}{k_t}}\bigg(\textrm{J}_0\left[\frac{k}{2k_t}\right]  - \left(\frac{k_t}{k} - i\right)\textrm{J}_1\left[\frac{k}{2k_t}\right]\bigg).
\end{split}
\end{equation}
These result in a relatively simple expression for the primordial power spectrum
\begin{equation}
\begin{split}
    \mathcal{P}_\mathcal{R}^{\textrm{(FIC)}}(k) = & \frac{3 k_t^2 k}{2 \varepsilon a_t^2 \pi^2}R^{(0)^2} \bigg(\textrm{J}_0\left[\frac{k}{2k_t}\right]\sin\left(\frac{k}{k_t}\right)\\
    & + \left(\cos\left(\frac{k}{k_t}\right) - \frac{k_t}{k} \sin\left(\frac{k}{k_t}\right)\right)\textrm{J}_1\left[\frac{k}{2k_t}\right] \bigg)^2.
    \label{eqn:fic_pps_og}
\end{split}
\end{equation}

\begin{figure}[t]
\includegraphics{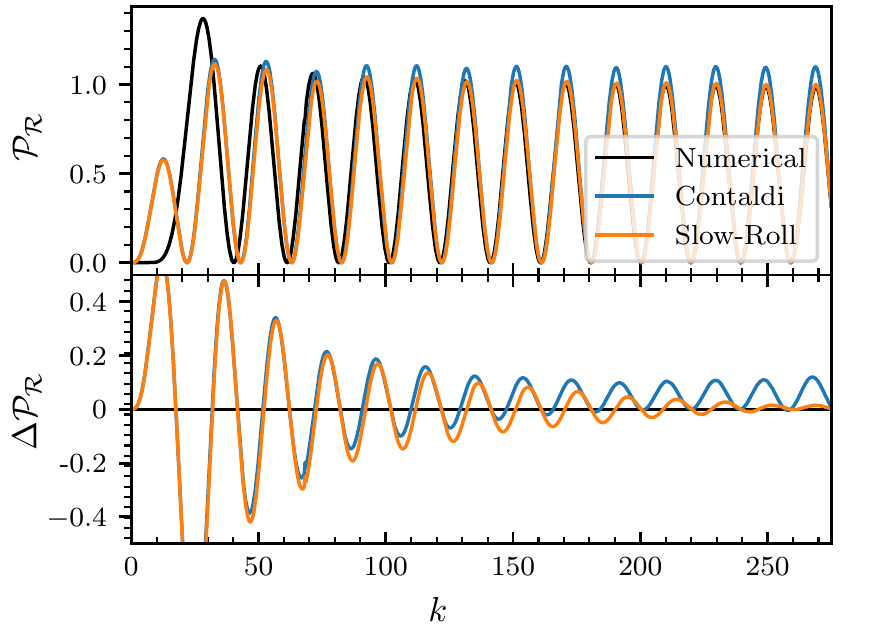}
\caption{\label{fig:pps_fic} Comparison of numerical evaluation and two approximations for the primordial power spectra produced by frozen initial conditions. Numerical evaluation took the scalar field potential as $V(\phi) = \frac{1}{2} m^2 \phi^2$ with $m = 1$ and was calculated using \texttt{oscode}~\cite{Agocs_2020}. For the approximations the free parameters were taken to be best fit values of $k_t = 9.34$ and $\nu_s = 1.52$.  The Contaldi method correctly predicts high $k$ oscillations but fails to account for the decay in peak height, while the slow-roll method appears to fit very well at high $k$. Both approximations at low $k$ correctly predict a cutoffs but underestimates the value at which it occurs.}
\end{figure}

The primordial power spectrum derived for FIC using the Contaldi approximation is shown in \cref{fig:pps_fic}. The spectrum is strikingly different to that found for the four other types of initial conditions. It still shows the same low $k$ cutoff, but now at high $k$ instead of plateauing it displays constant wavelength amplitude oscillations that go down to zero. Between these two regimes is an intermediate region of oscillations with varying heights. These features were also seen in the numerical investigation of FIC by \citet{Haddadin_2018b}.

By taking the the large $k$ limit of the analytic expression in \cref{eqn:fic_pps_og} we can quantify the oscillations 
\begin{equation}
    \lim_{k \gg k_t} \mathcal{P}_\mathcal{R}(k) = \frac{3k_t^3}{\varepsilon a_t^2 \pi^3}R^{(0)^2} \left(1 - \sin\left(\frac{3k}{k_t}\right)\right),
    \label{eqn:pps_fic_high_k_lim}
\end{equation}
finding they are sinusoidal with a constant wavelength of $\frac{2\pi k_t}{3}$. The FIC primordial power spectrum therefore has uniformly spaced peaks which suggests a similarity to the quantized primordial power spectra considered in~\cite{Lasenby_2020, Bartlett_2020}. The pseudo-quantized primordial power spectrum of FIC may therefore provide an improved fit to the Planck observations as was found by \citet{Bartlett_2020} for a fully quantized spectra.

However, as previously stated these primordial power spectra are just approximations. To get a handle on how good the approximation is we can also solve the Mukhanov-Sasaki equations numerically and compare. For the numerical solution we consider a scalar field with $V(\phi) = \frac{1}{2} m^2 \phi^2$ and use \texttt{oscode}~\cite{Agocs_2020} to evaluate the Mukhanov-Sasaski equations. The resulting $\mathcal{P}_\mathcal{R}$ is also shown in \cref{fig:pps_fic}. The Contaldi approximation appears to fit well at high $k$, except for not predicting the decay of the amplitude of the oscillations seen in the numerical result.  However around the low $k$ cutoff the approximation is much poorer, significantly underestimating the position of the cutoff.

\subsection{\label{ssec:apps_mdse}High $k$ Power Law Decay}

The approximation described in \cref{ssec:apps_meth} can be improved upon by changing the model of the inflationary epoch to slow-roll inflation rather than pure de Sitter inflation~\cite{Baumann_2009,Haddadin_2018b}. Henceforth referred to as the slow-roll approximation.  

Changing the inflationary stage to slow-roll introduces a single additional parameter $\nu_s$ that encapsulates how $H$ decreases during inflation. $\nu_s$ can be derived from the potential's slow-roll parameters, $\nu_s = \frac{3}{2}\sqrt{1 + 4\epsilon_V - \frac{4}{3}\eta_V}$, and can be related to the spectral index, $\nu_s = 2 - \frac{1}{2}n_s$~\cite{Baumann_2009}, when applied to \LCDM.  

For slow-roll inflation the solution to the Mukhanov-Sasaki equations has the general form~\cite{Baumann_2009,Haddadin_2018b}
\begin{multline} 
    v_k(\eta) = \sqrt{\eta - \frac{1}{k_t}}\bigg(E_k \textrm{H}^{(1)}_{\nu_s}\left[k\eta - \frac{k}{k_t}\right] 
    \\ + F_k \textrm{H}^{(2)}_{\nu_s}\left[k\eta - \frac{k}{k_t}\right] \bigg).
    \label{eqn:vk_mdS}
\end{multline}
With the resulting primordial power spectrum for such a solution is
\begin{equation} 
    \mathcal{P}_\mathcal{R}(k) \propto \frac{2^{2\nu_s}\Gamma(\nu_s)^2}{\pi^2}\frac{k^{3-2\nu_s}}{2\pi^2}\left|E_k - F_k\right|^2.
    \label{eqn:general_pps_mds}
\end{equation}

Using the slow-roll approach with our various initial conditions leads to complex and computationally expensive approximations for the primordial power spectrum. The slow-roll approximation of the FIC primordial power spectrum is also depicted in \cref{fig:pps_fic}. Comparing to the Contaldi method we see that the main difference is that under the new approximation the peak heights decay away as $k$ increases. As a result the slow-roll method seems to give a good fit to the numerically evaluated spectrum at high $k$. However at low $k$, like the Contaldi method, it significantly underestimates the position of the cutoff. 

To demonstrate this high $k$ decay of the primordial power spectrum is a general phenomena let us consider the Contaldi and slow-roll approximations when $k \gg k_t$. In the $k \gg k_t$ limit the general slow-roll solution (\cref{eqn:vk_mdS}) becomes
\begin{multline} 
    v_k(\eta) = \sqrt{\frac{1}{k}}\bigg(E_k \exp\left(i\left(k\left(\eta - \frac{1}{k_t}\right) - \frac{\nu_s \pi}{2} - \frac{\pi}{4}\right)\right) 
    \\ + F_k \exp\left(-i\left(k\left(\eta - \frac{1}{k_t}\right) - \frac{\nu_s \pi}{2} - \frac{\pi}{4}\right)\right) + O\left(\frac{k_t}{k}\right)  \bigg).
\end{multline}
Similarly the solution for pure de Sitter inflation \cref{eqn:vk_dS} becomes
\begin{multline} 
    v_k(\eta) = C_k e^{-i(k\eta - \frac{k}{k_t})}+ D_k e^{+i(k\eta - \frac{k}{k_t})} +  O\left(\frac{k_t}{k}\right). 
\end{multline}
These are equivalent if we make the identification
\begin{equation}
\begin{split}
    &C_k= \sqrt{\frac{1}{k}}F_k \exp\left(i\frac{\pi}{2}\left(\nu_s - \frac{3}{2}\right)\right), \\
    &D_k = \sqrt{\frac{1}{k}}E_k \exp\left(-i\frac{\pi}{2}\left(\nu_s - \frac{3}{2}\right)\right).
\end{split}
\end{equation}

Substituting these relations into the slow-roll formula for the primordial power spectrum \cref{eqn:general_pps_mds} gives
\begin{equation} 
    \mathcal{P}_\mathcal{R}^{\textrm{sr}}(k) \propto k^{4-2\nu_s}  \left|C_k e^{i\frac{\pi}{2}(\nu_s - \frac{3}{2})} - D_k e^{-i\frac{\pi}{2}(\nu_s - \frac{3}{2})}\right|^2.
    \label{eqn:pps_relation_start}
\end{equation}
Hence providing $\nu_s - \frac{3}{2}$ is small then in the high $k$ ($k \gg k_t$) regime the slow-roll inflation approximation and Contaldi approximation of the primordial power spectra are simply related by multiplication by a power law
\begin{equation}
    \mathcal{P}_\mathcal{R}^{\textrm{sr}}(k) \propto k^{3-2\nu_s}  \mathcal{P}_\mathcal{R}^{\textrm{Con}}(k).
    \label{eqn:pps_relation}
\end{equation}

Furthermore providing $C_k$ and $D_k$ are not the same order in $k /k_t$ in the $k \gg k_t$ limit then the above result extends to all values of $\nu_s$. This is in fact the case for the $C_k$ and $D_k$ results found for BD, HD and RSET as well as those found by \citet{Contaldi_2003} for RHM initial conditions.  

In \LCDM the primordial power spectrum is taken to be a pure power-law
\begin{equation}
    \mathcal{P}_\mathcal{R}^{\Lambda\textrm{CDM}}(k) = A_s \left(\frac{k}{k_*}\right)^{n_s - 1},
    \label{eqn:pps_lcdm}
\end{equation}
where $k_*$ is the pivot scale defined to be $0.05$ Mpc$^{-1}$. The Planck observations found $A_s = 2.092  \pm 0.034 \times 10^{-9}$ and $n_s = 0.9626 \pm 0.0057$ as the best fit values of these parameters~\cite{PLANCK_2018_VI}. Hence the observations support a weak power law decay of the primordial power spectrum. The best fit $n_s$ would correspond to $\nu_s = 1.5187 \pm 0.0029$. Observations therefore suggest $\nu_s - \frac{3}{2}$ is indeed small and hence for large $k$ \cref{eqn:pps_relation} should be a reasonable approximation for all initial conditions.

\begin{figure}[t]
\includegraphics{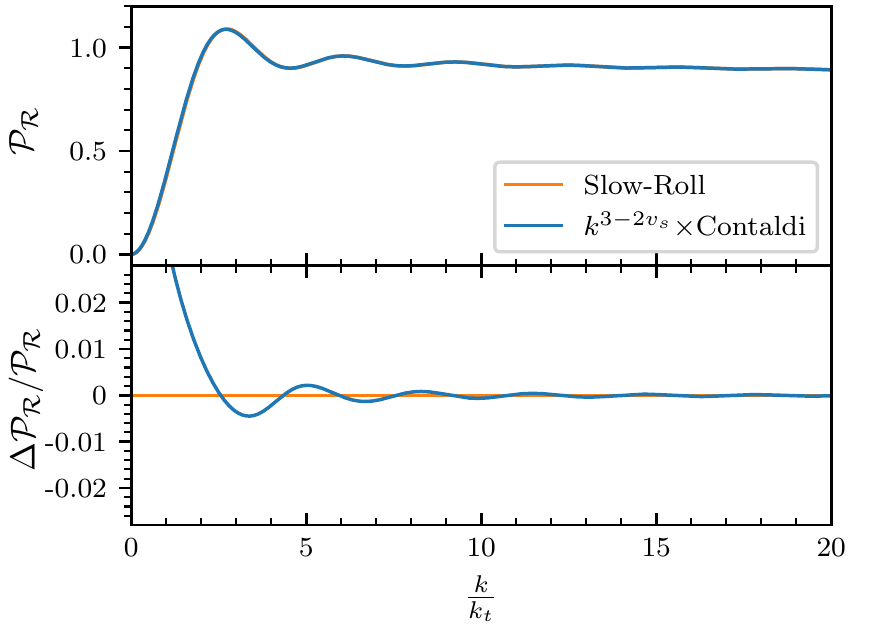}
\caption{\label{fig:pps_fic_comp_2} Primordial power spectrum for BD initial conditions using the slow-roll approximation and power-law times Contaldi approximation. In both $\nu_s$ was taken to be $1.5187$. The two different approximations are in good agreement at high $k$. The error only exceeds $1\%$ below the low $k$ cutoff.}
\end{figure}

As we shall go on to perform a Bayesian analysis of these models in \cref{sec:bayes} we need to keep the computational cost of calculating the primordial power spectrum low. As the Contaldi approximation primordial power spectra are much cheaper to compute then the slow-roll approximation primordial power spectra we shall make the further assumption that \cref{eqn:pps_relation} holds for all $k$, rather than just high $k$. \Cref{fig:pps_fic_comp_2} shows the fractional difference between the primordial power spectrum resulting from the two methods when applied to BD initial conditions. The figure shows good agreement between the two approaches in the intermediate and high $k$ range. For lower $k$ values, below the cutoff, there is a greater discrepancy, but as aforementioned for the Planck favoured $k_t$ values this regime corresponds to scales that are currently too large to be observed. Hence using \cref{eqn:pps_relation} for all $k$ should have no significant impact on observable predictions.

Using this assumption we derive our final forms for the primordial power spectra resulting from each of the initial conditions. For BD, HD, RSET and RHM we can most easily express these implicitly via
\begin{equation}
\begin{split}
    \mathcal{P}_\mathcal{R}(k) = 2 A_s \bigg(\frac{k^{n_s}}{k_*^{n_s-1}}\bigg)\left|C_k - D_k\right|^2.
\end{split}    
\end{equation}
With the $C_k$ and $D_k$ defined in \cref{eqn:cd_rhm,eqn:cd_bd,eqn:cd_hd,eqn:cd_rset}. For FIC a somewhat compact expression can be found
\begin{equation}
\begin{split}
    \mathcal{P}_\mathcal{R}^{\textrm{(FIC)}}&(k) =  \frac{\pi^2}{3}A_s\left(\frac{k_*}{\Delta k}\right)\left(\frac{k}{k_*}\right)^{n_s}\bigg(\textrm{J}_0\left[\frac{\pi k}{3 \Delta k}\right]\sin\left(\frac{2\pi k}{3 \Delta k}\right)\\
    & + \left(\cos\left(\frac{2\pi k}{3 \Delta k}\right) - \frac{3 \Delta k}{2 \pi k} \sin\left(\frac{2\pi k}{3 \Delta k}\right)\right)\textrm{J}_1\left[\frac{\pi k}{3 \Delta k}\right] \bigg)^2.
\end{split}
\end{equation}
Here $n_s$ and $A_s$ are defined so that the weak power law decays present at high $k$ in the BD, HD, RSET and RHM primordial power spectra are of the same form as the spectrum assumed in \LCDM, \cref{eqn:pps_lcdm}. While for FIC they are defined so that the cycle averages of the oscillations is also \cref{eqn:pps_lcdm}. $k_t$, or $\Delta k = \frac{2\pi k_t}{3}$ for FIC, is the additional free parameter in the primordial power spectra, $k_t$ having the same meaning as before and $\Delta k$ being the wavelength of the oscillations in the FIC spectra.

As BD, HD, RSET and RHM initial conditions all have power law decays in their primordial power spectra at high $k$ we expect that at high $k$ the four vacuum initial conditions would be indistinguishable from \LCDM. However at low $k$ the vacuum initial conditions have cutoffs not present in the \LCDM spectrum. As a result they may be distinguishable from \LCDM and from one another via observables that depend upon the low $k$ part of the primordial power spectrum.

On the other hand from \cref{fig:pps_fic} we can see that for FIC the primordial power spectrum never tends to a pure power-law decay. This suggests FIC may be more easily distinguished from \LCDM experimentally than the other initial conditions as it could be distinguished using observables that depend on any portion of the primordial power spectrum.

Using the above formulae for the primordial power spectra we can now determine the resulting CMB spectra for each type of initial conditions.

\section{\label{sec:obs}Observable Consequences}
\subsection{\label{ssec:oc_meth}Methodology}
For each set of initial conditions we shall consider the resulting TT, TE, and EE CMB spectra, which we compute using \texttt{CLASS}~\cite{CLASS_I,CLASS_II}. These spectra have been measured by the Planck satellite~\cite{PLANCK_2018_V}, with the results made publicly available on the Planck Legacy Archive~\cite{PLA}.

To calculate the CMB spectra, \texttt{CLASS} requires values for various cosmological parameters and the specification of the primordial power spectrum. In this section all cosmological parameters are taken to be the best fit parameters from the Planck 2018 results~\cite{PLA} except those controlling the primordial power spectrum. For the primordial power spectrum we provide \texttt{CLASS} with an external function which returns the appropriate approximate primordial power spectrum.

\subsection{\label{ssec:oc_results_QIC}Vacuum Initial Conditions}
\begin{figure*}[p]
\includegraphics{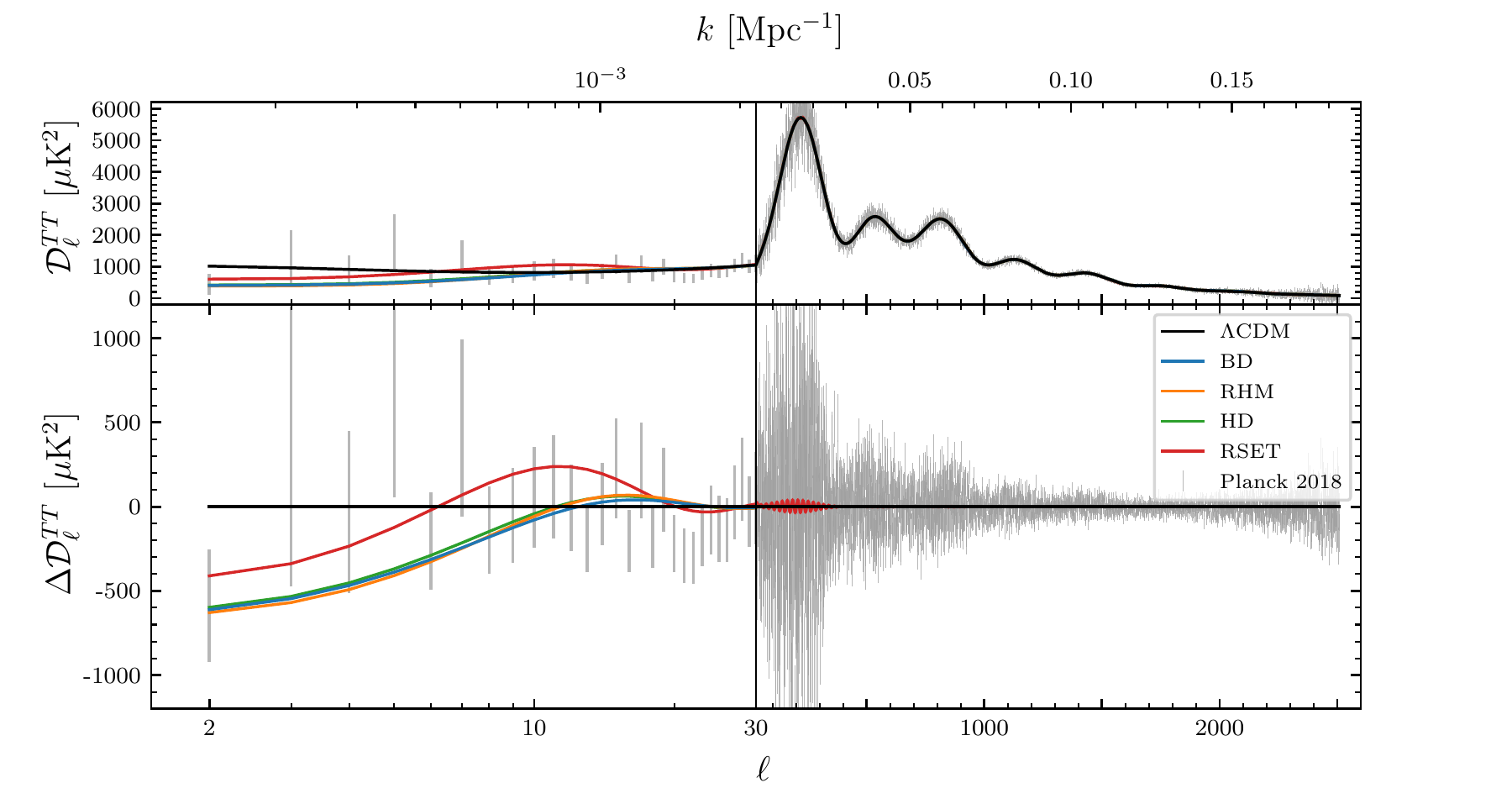}
\caption{\label{fig:dl_qic_tt} CMB temperature spectrum for BD, HD, RHM, and RSET initial conditions along with the best fit \LCDM model and experimental data from Planck 2018. The free parameter $k_t$ is taken to be $5 \times 10^{-4}$ Mpc$^{-1}$, which is chosen so that the cutoffs in the primordial power spectra were in the same region as the best fit cutoffs found in~\cite{Contaldi_2003}. At high $\ell$ there is no noticeable difference between any of the initial conditions and \LCDM. However at low $\ell$ there is some variation between \LCDM, RSET and the other three. All four initial conditions lead to a suppression of the low multipoles compared to \LCDM which may make them favoured over the baseline \LCDM model. The top axis gives an approximate correspondence between $k$ and $\ell$ using the Limber approximation~\cite{Handley_2019}.}

\includegraphics{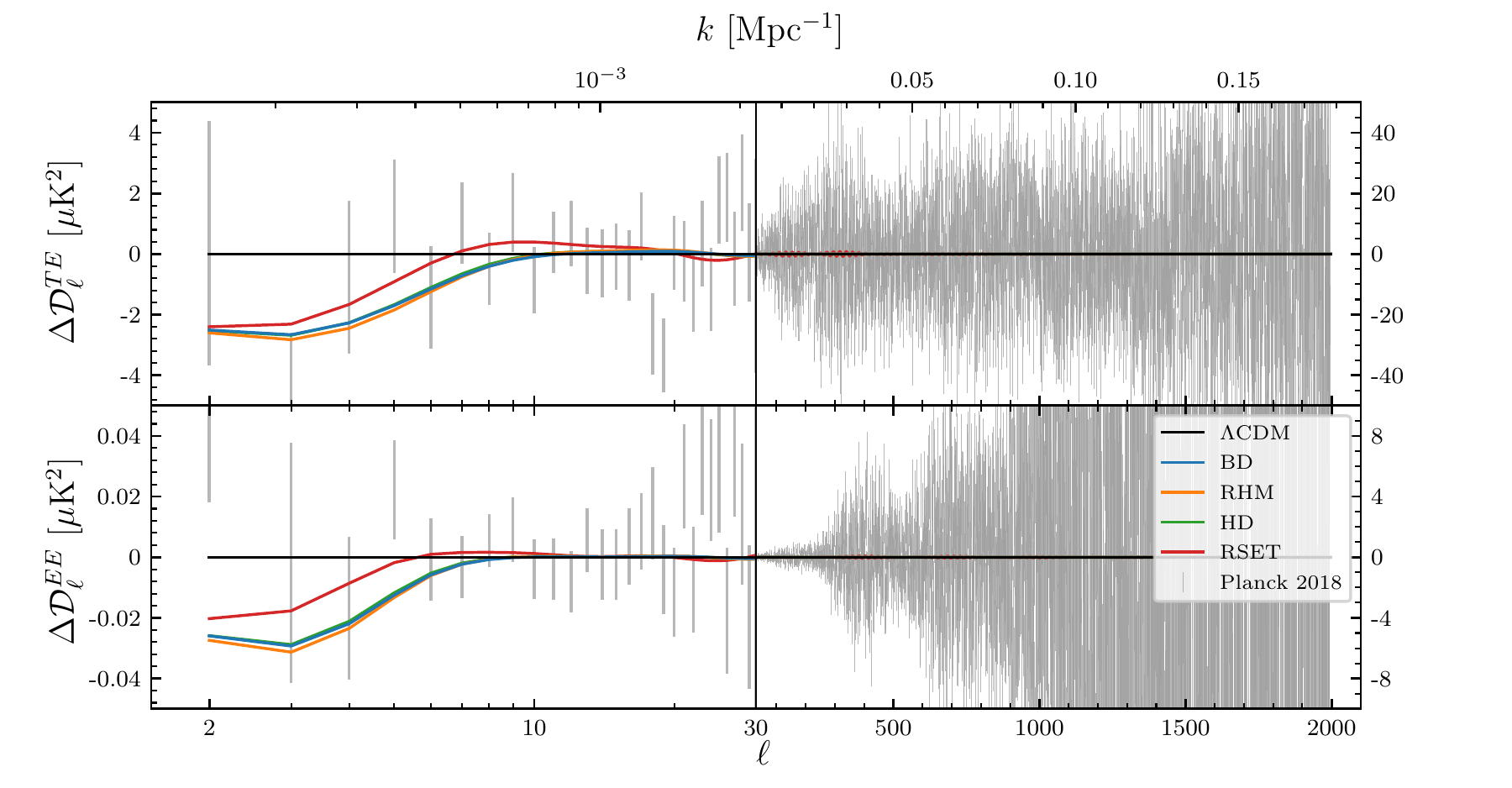}
\caption{\label{fig:dl_qic_te_ee} CMB TE and EE spectra for BD, HD, RHM, and RSET initial conditions along with experimental data from Planck 2018. Spectra presented as differences from the Planck 2018 best fit \LCDM model. Note that unlike Figure 7 the right hand portion of the graphs have different scales to their left hand portion as there is such a large magnitude difference. As above the free parameter $k_t$ is taken to be $5 \times 10^{-4}$ Mpc$^{-1}$. Similarly to the TT spectra, for all vacuum initial conditions both the TE and EE spectra are indistinguishable from \LCDM at high $\ell$ and have a relative suppression of power at low multipoles, with noticeable differences between RSET and the other initial conditions spectra at low $\ell$. }
\end{figure*}

The CMB temperature spectra for BD, RHM, HD, and RSET are shown in \cref{fig:dl_qic_tt} with \mbox{$k_t = 5 \times 10^{-4}$ Mpc$^{-1}$}. For larger $\ell$ all four models are indistinguishable from \LCDM. This is not unexpected as at high $k$ the primordial power spectra of these initial conditions tends to the power-law spectra of \LCDM. Furthermore, as for BD, RHM, and HD the low $k$ primordial power spectra were very similar the corresponding $\mathcal{D}^{TT}_\ell$ are also almost identical at low $\ell$ as well, with a suppression in $\mathcal{D}^{TT}_\ell$ at low multipole moments compared to \LCDM. The differences between these three initial conditions are far smaller than the uncertainties of the Planck 2018 data and so they are evidently not observationally distinguishable from one another currently, although they are possibly distinguishable from \LCDM. RSET differs noticeably from the other three initial conditions at low $\ell$, having a smaller low multipole suppression and a peak not present in the others.

\begin{figure*}[p]
\includegraphics{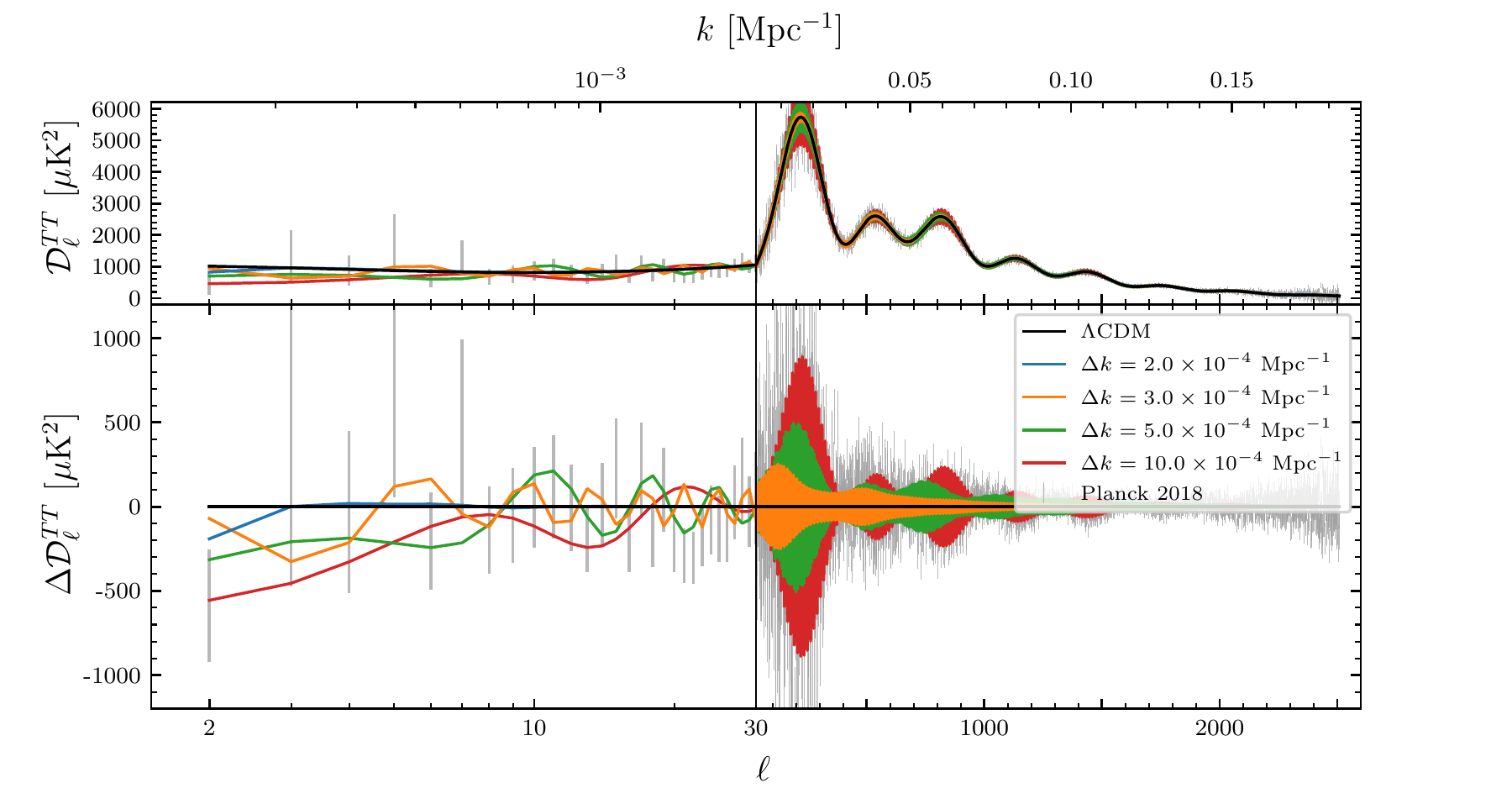}
\caption{\label{fig:dl_fic} Unlensed CMB temperature spectrum produced by frozen initial conditions for various values of the $\Delta k$ parameter along with the best fit \LCDM model and experimental data from Planck 2018. For larger $\Delta k$ the FIC spectra show noticeable oscillations about the \LCDM spectrum. However for small $\Delta k$ the oscillations seem to average out producing a spectra very similar, or even identical, to \LCDM. The oscillations for some intermediate values of $\Delta k$, for example $5.0 \times 10^{-4}$ Mpc$^{-1}$, may provide a potential explanation for the drop in power seen at around $\ell = 20$.}

\includegraphics{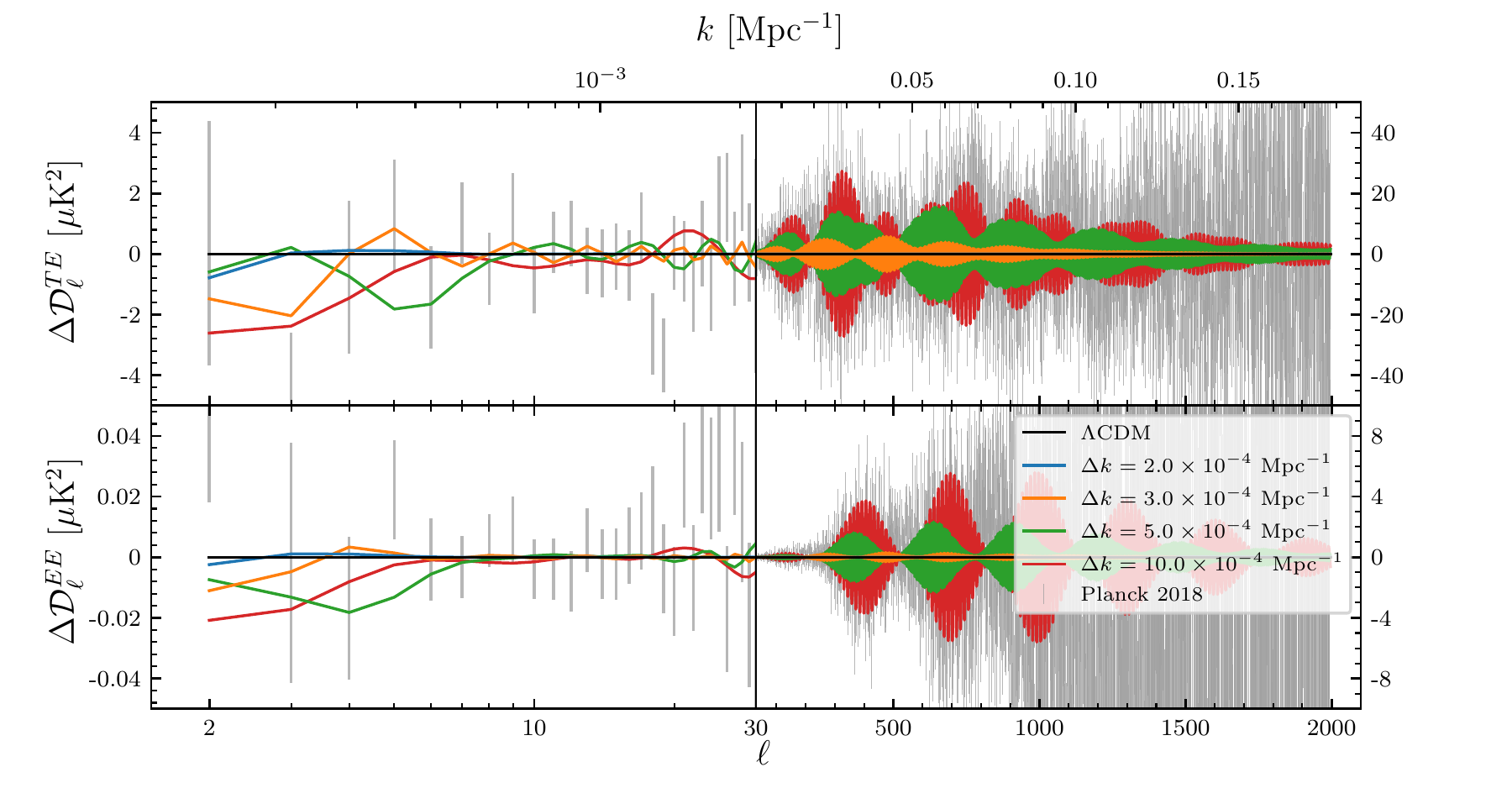}
\caption{\label{fig:dl_fic_TE_EE} Unlensed CMB TE and EE spectra for frozen initial conditions with various values of the $\Delta k$ parameter along with experimental data from Planck 2018. Spectra presented as differences from the Planck 2018 best fit \LCDM model. The right and left hand portion of the graphs have different scales due to there being a large magnitude difference. As was seen in the TT spectra the FIC spectra vary strongly with the value of $\Delta k$, with higher values displaying large oscillations and low values being near indistinguishable from \LCDM. }
\end{figure*}

For higher $k_t$ the difference between the vacuum initial condition spectra and that of \LCDM are more exaggerated, with greater low multipole suppression and more prominent oscillations. Conversely for lower $k_t$ the differences are smaller with the predicted spectra all being identical at $k_t = 0$, which follows from the fact in this limit the primordial power spectra all converge to the power law assumed in \LCDM.  

Similar behaviour is seen in both the TE and EE spectra of the vacuum initial conditions, as shown in \cref{fig:dl_qic_te_ee}. Together the three spectra all suggest that it may be both possible to distinguish a RSET vacuum from other potential vacua via observations of the CMB spectra at low $\ell$. In addition some, or all, of the vacua may provide better fits to current observations than \LCDM as they display a low $\ell$ power suppression, not predicted by \LCDM, which was found in \cite{Contaldi_2003} to give an improved fit to the WMAP observations. In \cref{sec:bayes} we shall use the Planck likelihoods to see if there is currently any significant evidence in favour of any of the vacuum initial conditions over the others and \LCDM.

\subsection{\label{ssec:oc_results_FIC}Frozen Initial Conditions}

Let us consider the unlensed temperature spectrum resulting from FIC shown in \cref{fig:dl_fic}. For $\Delta k$ above $2.4 \times 10^{-4}$ Mpc$^{-1}$ the oscillations in the primordial power spectrum lead to noticeable oscillations in the CMB spectra. As $\Delta k$ increases the oscillation wavelength increases and any averaging effect due to each $\ell$ corresponding to a range of $k$ diminishes, leading to larger and larger oscillations about the \LCDM spectra. For \mbox{$\Delta k < 15.0 \times 10^{-4}$ Mpc$^{-1}$} the oscillations still remain within the uncertainties of the Planck data, however for larger $\Delta k$ they exceeded the experimental errors, suggesting very high $\Delta k$ values are not consistent with observation.

Conversely for small $\Delta k$ the oscillations in the primordial power spectrum seem to be too fine to be seen by the resolution the multipole moments of the CMB spectra provide, and so they average out leaving a temperature spectrum that is nearly identical to \LCDM. Providing $\Delta k$ is small but still greater than \mbox{$1.0 \times 10^{-4}$ Mpc$^{-1}$} there remains some small differences between FIC and \LCDM with FIC showing a slight suppression of low multipoles due to its primordial power spectrum cutoff. However, below $\Delta k = 1.0 \times 10^{-4}$ Mpc$^{-1}$ this suppression also disappears resulting in FIC predicting an identical spectra to \LCDM. This result demonstrates that the universe starting in a quantum vacuum state is not actually a requirement to produce unlensed CMB spectra similar to those observed by Planck. 

In addition FIC can provide a potential explanation for two of the unexplained features in the CMB spectra, the suppression of power at low multipoles and the dip in power around $\ell=20$, this can be seen in the figure for $\Delta k = 5.0 \times 10^{-4}$ Mpc$^{-1}$. Hence it qualitatively appears for some $\Delta k$ values the FIC CMB spectra could match experimental observations better than \LCDM. However, the oscillations at high $\ell$ may lead to a poorer fit at high multipoles. 

Similar features are also seen in the FIC unlensed TE and EE CMB spectra, \cref{fig:dl_fic_TE_EE}. Again higher $\Delta k$ valued FIC models show significant oscillations about the \LCDM spectrum, which for $\Delta k < 15.0 \times 10^{-4}$ Mpc$^{-1}$ remain smaller than the uncertainties on the experimental data. In addition for low $\Delta k$ the FIC TE and EE spectra are also near indistinguishable from \LCDM.

Hence the unlensed FIC spectra suggest FIC could either: provide an alternative theoretic paradigm for understanding results currently understood to stem from an inflated quantum vacuum; or better fit the Planck observations by simultaneously having a low multipole power suppression and fitting the $\ell \approx 20$ power dip.

Unfortunately even with the highest accuracy settings available to us \texttt{CLASS} was unable to calculate lensed CMB spectra for FIC. Further investigation suggests that the program cannot currently compute lensing for primordial power spectra with fine feature at high $k$ as are present for FIC. To modify the code to be able to handle such features is beyond the scope of this paper. Consequently our analysis of FIC observables is currently limited to unlensed spectra. Furthermore as the Planck likelihoods require lensed spectra as inputs this prevents us performing a Bayesian analysis on FIC at this time. The required modification to \texttt{CLASS} and subsequent further investigation of FIC is left to a future work.

\section{\label{sec:bayes}Bayesian Model Comparison}

\subsection{\label{ssec:bayes_meth}Methodology}

In the previous section there were qualitative suggestions that the effects of some quantum vacuum derived initial conditions may be observationally distinguishable from other as well as from \LCDM. In addition such models may provide potentially better fits to existing CMB observations than \LCDM due to their low power suppression. To determine quantitatively if current observations favour certain quantum vacuums, or \LCDM, we can compute Bayes' factor~\cite{Mackay_2003}. Using Bayes' factor to compare models has the advantage of automatically including a ``Occam penalty'' wherein models with more parameters are punished and hence helps avoid overfitting, which in this case is a risk as the vacuum initial condition models introduce an additional parameter ($k_t$) not present in \LCDM. 

To calculate these Bayes factors we use \texttt{Cobaya}~\cite{Cobaya} a code designed for cosmological Bayesian analyses, which we run on the CSD3 high performance computing facility~\cite{HPC}. Within \texttt{Cobaya} we employed the theory code \texttt{CLASS}~\cite{CLASS_I,CLASS_II}, and the sampler \texttt{PolyChord}~\cite{Polychord_1,Polychord_2} due to us needing to compute the Bayesian evidences to calculate Bayes' factor~\cite{Mackay_2003}. The data set utilized to compare the models is the Planck TTTEEE+lowE+lensing likelihoods~\cite{PLANCK_2018_V,PLANCK_2018_VIII,PLA}. For all the cosmological parameters we imposed uniform priors, detailed in \cref{tab:prior}. 

\renewcommand{\arraystretch}{1.2}
\begin{table}[t]
\caption{\label{tab:prior}%
Prior distributions for the cosmological parameters of our models, all of which are taken to be uniform. The first $6$ parameters are the standard \LCDM parameters and are needed by all models, whereas $k_t$ is only required by the four vacuum initial conditions models. }
\begin{ruledtabular}
\begin{tabular}{ccc}
\textrm{Parameter}&
{\textrm{Prior Minimum}} & {\textrm{Prior Maximum}}\\
\colrule
$\log{10^{10}A_s}$ & $2.2$ & $3.5$ \\
$n_s$ &  $0.885$& $1.04$\\
$\Omega_b h^2$ & $0.019$ & $0.025$\\
$\Omega_{c} h^2$ & $0.095$ & $0.145$\\
$100 \theta_s$ & $1.03$ &  $ 1.05$ \\
$\tau_{\textrm{reio}}$ & $0.01$ &  $ 0.10$ \\
\colrule
$k_t$ & $0$  & $50.0 \times 10^{-4}$ Mpc$^{-1}$ \\
\end{tabular}
\end{ruledtabular}
\end{table}

\subsection{\label{ssec:bayes_results}Results and Discussion}

\begin{figure*}
\includegraphics{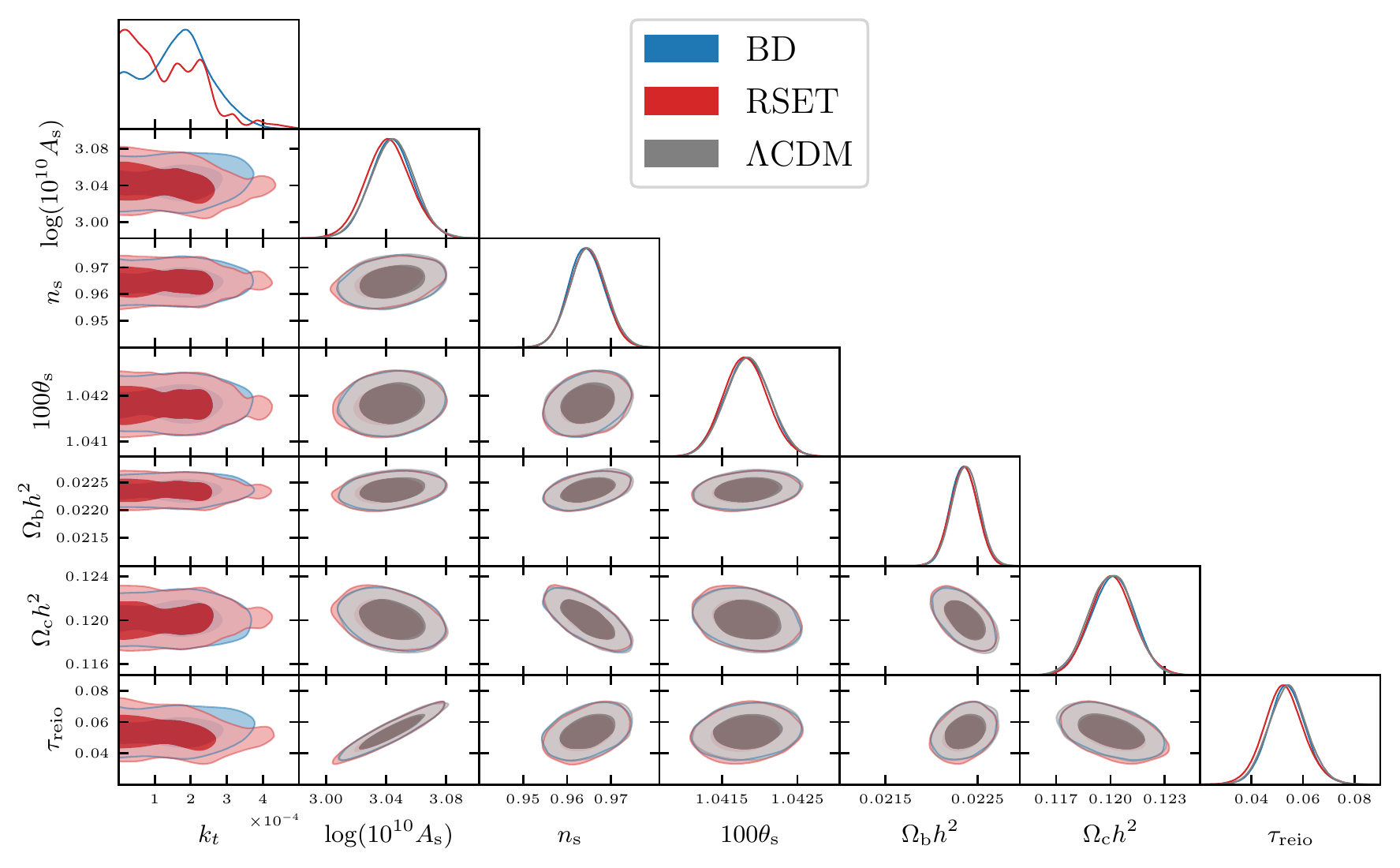}
\caption{\label{fig:triangle_plot}Posterior distributions of the parameters of \LCDM, BD, and RSET models. RHM and HD posterior distributions were found to be near identical to the BD posterior shown here. For the $6$ shared parameter the posteriors are all very similar and in good agreement with the Planck 2018 results for \LCDM. The marginalized $k_t$ posterior of BD shows a noticeable peak around $2.0 \times 10^{-4}$ Mpc$^{-1}$ whereas for RSET the $k_t$ posterior is maximum at $0$ and generally decays as $k_t$ increases. Given the predictions of BD, RSET and \LCDM are equivalent when $k_t = 0$ this suggests the BD model can provide a better fit to the Planck data than \LCDM. In addition it would seem \LCDM is a better model then RSET on grounds of being more parsimonious. Plot produced using \texttt{GetDist}~\cite{GetDist}.}
\end{figure*}

\cref{fig:triangle_plot} shows the resulting posterior distributions for the BD, RSET, and \LCDM models. RHM and HD posteriors are not depicted to avoid duplication, as they were indistinguishable to that of BD, consequently all conclusions we draw for BD apply to RHM and HD as well. In all three shown posteriors the marginalized likelihoods for the six baseline \LCDM parameters are very similar and in good agreement with the 2018 findings of the Planck collaboration~\cite{PLANCK_2018_VI}.

\begin{figure}
\includegraphics{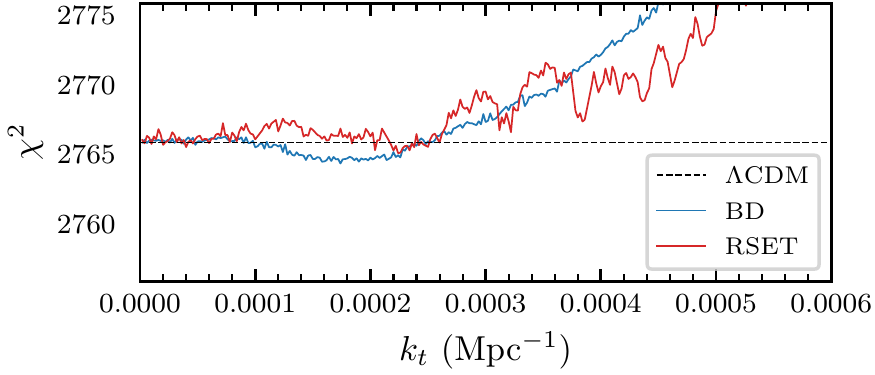}
\caption{\label{fig:chi_sq}Profile likelihood of BD and RSET models over the parameter $k_t$, with the $\chi^2$ for the best fit \LCDM model shown for comparison. At each fixed value of $k_t$ a Nelder-Mead optimization is performed over the remaining parameters to find the best fit BD/RSET model with that value of $k_t$. From this model we then calculate $\chi^2 = -2 \ln(\textrm{P}(\textrm{Data} | \textrm{Model}))$. We find BD shows a significant range of $k_t$ values where it can provide a better fit to the Planck 2018 data than \LCDM. Conversely RSET only improves on the best fit \LCDM in a very narrow region of parameter space and only by a small amount.}
\end{figure}

However, the $k_t$ posterior shows very different behaviour for BD and RSET. At $k_t = 0$ the vacuum initial condition models are equivalent to \LCDM. As $k_t$ increases for BD the likelihood is found to increase to a peak around $2.0\times 10^{-4}$ Mpc$^{-1}$, suggesting that the BD model can provide a better fit to the Planck likelihoods than \LCDM. The profile likelihood plot in \cref{fig:chi_sq} confirms this conclusion, showing a range of $k_t$ values from $1.0$ to $2.4\times 10^{-4}$ Mpc$^{-1}$ where BD can provide a better fit to the Planck Data than \LCDM. This was not unexpected given that previous studies had found a low multipole power suppression relative to \LCDM, as is present in BD, provides better fits to CMB observations \cite{Contaldi_2003}. 

\begin{figure}
\includegraphics{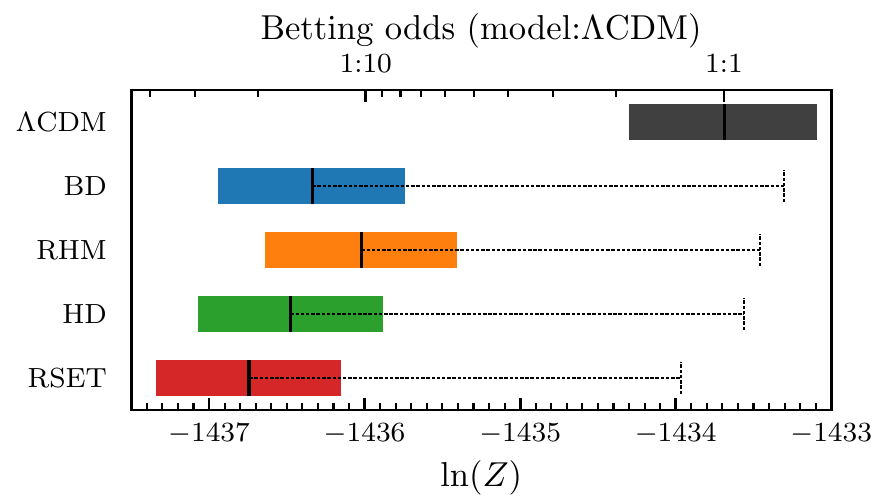}
\caption{\label{fig:bayes_evidence}Log Bayesian evidences of the vacuum initial condition and \LCDM models. For each model the central black line is the calculated log evidence and the coloured bar about it gives the $2\sigma$ confidence interval in that value. The dashed lines shows the relative Occam penalty incurred by the vacuum initial condition models, estimated from difference in KL divergence~\cite{2021arXiv210211511H} between the model and \LCDM, calculated using \texttt{anesthetic}~\cite{anesthetic}. The results show that \LCDM is significantly better supported by the observations than the vacuum initial condition models, primarily due to the Occam penalty.}
\end{figure}

Surprisingly the same is not true for RSET where as $k_t$ increases the $k_t$ marginalized likelihood generally decreases, never rising above the $k_t = 0$ value. In addition the profile likelihood for RSET only shows a small improvement over \LCDM for a small range of $k_t$. Hence the weaker low power suppression and oscillations in the CMB spectra induced in the RSET models seem not to be favoured by current observations. 

Therefore from the posteriors and profile likelihoods it seems that RSET is not favoured over \LCDM, but BD may be depending on the comparative sizes of the evidence gain from the improved fit and the Occam penalty. To test these observations we compute Bayes' ratio
\begin{equation}
    B = \frac{P(\textrm{Model 1}|\textrm{Data}) P(\textrm{Model 1})}{P(\textrm{Model 2}|\textrm{Data}) P(\textrm{Model 2})} = \frac{Z^\textrm{(1)}}{Z^\textrm{(2)}}.
    \label{eqn:Bayes_ratio}
\end{equation}
Where we have taken a priori the models to be equally likely. Here $Z^{(i)}$ is the Bayesian evidence of model $i$.

The Bayesian analysis we performed gave the evidences shown in \cref{fig:bayes_evidence}. From these results we immediately see that \LCDM is the best of the models at describing the Planck 2018 observations. Evidently the Occam penalty for introducing an additional parameter overwhelmed the slightly improved fit to the data that BD, RHM, and HD provided. Comparing \LCDM to the best supported of the quantum vacuum models, RHM, we compute a Bayes ratio with $2\sigma$ confidence interval of $10$ $[4,23]$. Showing strong support for \LCDM over even the best of the vacuum initial condition models. 

As expected due to producing very similar CMB spectra the BD, RHM, and HD evidence confidence intervals greatly overlap. Calculating Bayes' ratio from any pair of these models gives a confidence interval that includes 1, showing they are not observationally distinguishable using the 2018 Planck data. In addition while RSET did produce somewhat different predictions for its CMB spectra the Bayes' ratio between RSET and the other three produces $2\sigma$ confidence intervals including 1. As an example the ratio between RHM and RSET is $2.1$ with a $2\sigma$ confidence interval of $[0.9,5]$. We therefore conclude that the Planck 2018 observations are not sufficient to distinguish between any of the vacuum initial condition models.

\section{\label{sec:conc}Conclusions}
We have investigated the effects of setting various initial conditions on the universe's scalar perturbations before, rather than during, inflation. By adopting an approximate approach, based upon the work of \citet{Contaldi_2003}, we have been able to draw generic conclusions that disentangle the effects of initial conditions from those of the choice of the inflaton potential. 

For the four quantum vacuum initial conditions considered we found a suppression of the lower multipole moments in the CMB TT, TE, and EE spectra when compared to the spectrum predicted by \LCDM. This low multipole suppression present in the vacuum initial condition models in some cases provides a better fit to the Planck 2018 observations than the standard \LCDM model. However, using Bayes' factor we demonstrated \LCDM remains the model better supported by the data on grounds of it being more parsimonious. By showing the differences in the primordial power spectra and CMB spectra resulting from the different vacua we indicated that some are at least in theory distinguishable from others. A Bayesian analysis of the models demonstrated the Planck observations do not provide evidence to support one vacua over another, so to distinguish vacua would require further experimental results. 

In addition to vacua initial conditions we have also investigated Frozen Initial Conditions (FIC), which correspond to a white-noise initial state at the big-bang singularity. FIC produce a primordial power spectrum with a suppression at low comoving wavevectors and large oscillations at high $k$ whose amplitude decay as a weak power-law. Hence the primordial power spectrum for FIC differs from that of \LCDM for all $k$ and so in theory FIC may be more easily distinguishable from \LCDM than the other initial conditions would be, as it could be distinguished through observables depending on either the low $k$ or high $k$ features of the primordial power spectrum.

We then found the unlensed CMB spectra predicted by FIC, and demonstrated that it strongly depends on the wavelength of the oscillations in its primordial power spectrum, $\Delta k$. For large $\Delta k$ oscillations about the best fit \LCDM model were predicted at higher $\ell$, as well as a suppression of the lower multipoles. In some cases these oscillations provided a potential explanation for the dip in power seen in the CMB TT spectra around $\ell \approx 20$. While for for smaller $\Delta k$, less than $2.4 \times 10^{-4}$ Mpc$^{-1}$, it was found that the predictions of frozen initial conditions became near identical to those of \LCDM. The oscillations in the primordial power spectrum seemingly being averaged out by the limited resolution provided by discrete multipole moments. Due to this equivalence between the predictions of FIC and \LCDM, FIC may provide an alternative explanation for existing experimental results previously believed to be understood via \LCDM and its explanation of being caused by a quantum vacuum initial state. This observation may be relevant to bouncing cosmologies~\cite{Thavanesan_2021} where such a thermal white noise spectra may have potentially arisen during the collapsing epoch before the bounce.

We feel the results of this paper opens up several interesting directions for further study. Firstly the investigation of the effect of when during kinetic dominance vacuum initial conditions were set, here we only considered setting them at the end of kinetic dominance. Secondly whether the FIC and \LCDM equivalence for low $\Delta k$ continues to observables such as the lensed CMB spectra (that we found \texttt{CLASS} unable to calculate), LSS, and BAO. Finally whether FIC for intermediate $\Delta k$ provides an improved fit to the Planck observations by simultaneously encapsulating a low monopole power suppression and a dip in power around $\ell = 20$.

\section*{\label{sec:ack}Acknowledgments}
We wish to thank Lukas Hergt for his invaluable help and advice on using \texttt{CLASS} and \texttt{Cobaya}, as well as Ayngaran Thavanesan for many useful discussions. TGJ thanks the Cavendish Laboratory's part III project scheme and STFC for their support via grant number ST/V506606/1. WJH is grateful to have been supported by a Gonville \& Caius Research Fellowship and a Royal Society University Research Fellowship.

\bibliography{ICBI}%

\end{document}